\documentclass[aps]{revtex}

\usepackage[T1]{fontenc}
\usepackage[latin1]{inputenc}
\usepackage{a4}
\usepackage{epsfig}
\usepackage{graphics}
\usepackage{graphicx}

\def\i{\'\i}

\def\be{\begin{equation}}
\def\ee{\end{equation}}
\def\bea{\begin{eqnarray}}
\def\eea{\end{eqnarray}}

\def\<{\langle}
\def\>{\rangle}



\begin{document}

\title{ 
       Numerical Simulation of  $N$-vector Spin Models \\
       in a Magnetic Field
}

\author{Tereza Mendes and Attilio Cucchieri\footnote{Work presented by T.\ Mendes
at the {\it IV Brazilian Meeting on Simulational Physics} -- Ouro Preto, August 2005.}
}

\address{
Instituto de F\'\i sica de S\~ao Carlos -- 
Universidade de S\~ao Paulo \\
C. P. 369, 13560-970, S\~ao Carlos, SP, Brazil
}

\date{September 2005}

\maketitle

\begin{abstract}
Three-dimensional $N$-vector spin models may define universality classes for 
such diverse phenomena as i) the superfluid transition in liquid helium 
(currently investigated in the micro-gravity environment of the Space 
Shuttle) and ii) the transition from hadronic matter to a
quark-gluon plasma, studied in heavy-ion collisions at the laboratories
of Brookhaven and CERN.
The models have been extensively studied both by field-theoretical 
and by statistical mechanical methods, including Monte Carlo simulations 
using cluster algorithms. These algorithms are applicable also in the 
presence of a magnetic field.
Key quantities for the description of the transitions above --- 
such as universal critical amplitude ratios and the location of the
so-called pseudo-critical line --- can be obtained from the models' magnetic 
equation of state, which relates magnetization,
external magnetic field and temperature. Here we present an improved parametrization 
for the equation of state of the models, allowing a better fit to the numerical 
data. Our proposed form is inspired by perturbation theory, with coefficients
determined nonperturbatively from fits to the data.

\end{abstract}

\section{Introduction}
\noindent

The $N$-vector (continuous-spin) models are proposed as class representatives 
for phase transitions in several interesting physical systems,
such as the superfluid transition in liquid helium, in the $N = 2$ case
\cite{Chaikin},
and the deconfinement transition in quantum chromodynamics (QCD) with 2 
flavors of light quarks, in the $N = 4$ case.
The deconfinement phase transition is obtained when hadrons (e.g.\ protons and 
neutrons) melt into a quark-gluon plasma at very high temperatures, such as the
temperatures that were present at beginning of the universe. There is great interest in
describing this transition and in obtaining the properties of the high-temperature 
phase, a new state of matter that might be present today in the interior of neutron
stars.
In the case of two degenerate light quark flavors (i.e.\ {\em up} and {\em down}), 
the transition is believed to be described by the three-dimensional 4-vector model.
More precisely, one invokes the effective $\sigma$-model \cite{pw}, a 
(three-dimensional)
Ginzburg-Landau effective theory written assuming universality and respecting the 
chiral symmetry of QCD. The theory relates the QCD order parameter, which is the 
chiral condensate $<{\overline \psi \,\psi}>$ (where $\psi$ is the quark field),
to the magnetization of a continuous-spin model.
The analogue of the magnetic field $H$ is given by the (nonzero but small) quark mass
and the reduced 
temperature is defined for lattice QCD as ${t}\sim 6/g^2 - 6/{g_c}^2(0)$, where 
$g$ is the lattice coupling.
For two quark flavors one then obtains ---  if transition is second order ---
a three-dimensional 4-vector model in the presence of a magnetic field.
The equivalence just described allows one to study critical properties of the
QCD phase transition from the spin-model equation of state. 
One of these properties is the so-called pseudo-critical line, the analogue of the 
critical point for the case of nonzero magnetic field. The pseudo-critical line is 
defined by the points where the susceptibility shows a (finite) peak,
corresponding to the rounding of the divergence observed for $H=0, \,T=T_c$.

We note that the equivalence between 2-flavor lattice QCD and the 4-vector model
is still not verified in comparisons of the respective numerical data (see e.g.\ 
\cite{Mendes:2005yp,Mendes:2005wf}
and references therein). Thus, a better knowledge of the magnetic equation 
of state for the 4-vector model is of great importance to achieve higher precision
in this comparison, to verify if the equivalence really holds and/or to establish 
the nature of the QCD phase transition, recently claimed to be of first order
\cite{D'Elia:2005bv}.

Also in the case of the 2-vector (or $XY$) model, a high-precision
nonperturbative determination of the equation of state is of interest,
since there are still discrepancies between the latest experimental 
and perturbative-renormalization-group values for critical
quantities at the superfluid helium transition \cite{Lipa}.
Note that both these determinations are very accurate, while the
available nonperturbative values (from Monte Carlo simulations at zero 
magnetic field \cite{Hasenbusch,Cucchieri:2002hu}) are not as precise.

\vskip 3mm
The Hamiltonian for the $N$-vector models is given by
\begin{equation}
{\cal H}\;=\;-J \,\sum_{<i,j>} {\bf S}_i\cdot {\bf S}_j
         \;-\; {\bf H}\cdot\,\sum_{i} {\bf S}_i \,,
\end{equation}
where the spin variables ${\bf S}_i$ are taken as
vectors on a sphere of unit radius in an $N$-dimensional space.
The main difference with respect to the Ising case is the possibility
of configurations where the spins are locally aligned but for long
distances this alignment is lost, yielding a null average for the
magnetization. Such configurations --- called {\em spin waves} ---
possess arbitrarily low energy and tend to destroy the
order of the system even at low temperatures.
In $d=3$ the models display a phase transition, with the presence of
spontaneous magnetization below the critical temperature, but
the spin waves lead to Goldstone-mode induced singularities, causing
a diverging susceptibility
and strong finite-size effects for all $T<T_c$ when $H\to0$.

Monte Carlo simulations can be performed very efficiently for $N$-vector models,
via cluster algorithms, also in a magnetic field \cite{Engels:1999wf,Engels:2000xw,Engels:2001bq}.
The simulation 
in the presence of external field has the advantage that one can obtain 
the equation of state directly from the data, as described below. Also, 
in this case one can ``measure'' the actual magnetization of the system,
without the need of an estimator such as the absolute value. In fact,
the magnetic field already selects only one of the equivalent (zero-field) 
ground states that would lead to the same value of the estimator but would
average to zero over the simulation in the zero-field case.
The numerical simulation is done via cluster algorithm, which can be applied 
to the case of nonzero magnetic field by employing the ghost-spin trick.
The observables are the magnetization parallel to the magnetic field and
the two susceptibilities (parallel and orthogonal to $H$).

We have recently proposed \cite{Cucchieri:2004xg}
an improved parametric form for the equation of state of 
the models. Our proposed form --- inspired by perturbation theory --- 
is a series expansion with two sets of terms, which contribute   
(mainly) separately to the description of the high- and low-temperature
regions of the phase diagram. In this way we achieve a better description 
of the low-temperature phase at zero magnetic field (i.e.\
the coexistence line), characterized by the singularities described above.
As a consequence, we are able to obtain a very precise
characterization of the pseudo-critical line for the 4-vector model.
We are currently applying the parametrization to a study of the $N=2$ case.
This will allow a better determination of the ratio of critical amplitudes 
for the specific heat in the superfluid helium transition.
Here we present preliminary results of this study, comparing the 
equations of state obtained for the cases $N =$ 2, 4.

\section{The magnetic equation of state}
At infinite volume, the scaling function for the singular part of the free
energy is given by
\begin{equation}
F_s(t,h) \;=\;
b^{-d}\,F_s(b^{y_t} t,b^{y_h}{h})\,,
\end{equation}
where $b$ is arbitrary, $ t = (T-T_c)/T_0$, ${h} = H/H_0$
and $y_t$, $y_h$ are related to the usual critical exponents $\beta$ and $\gamma$.
The above form implies the relation between magnetization and the applied 
magnetic field, known as the magnetic equation of state
\begin{equation}
M/h^{1/{\delta}} =
f_M({t}/{h}^{1/{\beta} {\delta}})\,.
\end{equation}
Equivalently
\begin{equation}
y\;=\;f(x) \,,
\end{equation}
with $y={h}/M^{{\delta}}$,
$x={t}^{{\beta}}/M$. The normalization constants are given by
$ f(0)\,=\,1$, $f(-1)\,=\,0$.
In the case of the $N$-vector models, the singularities at low temperature
determine the behavior of the magnetization as the square root of $H$.
This behavior (i.e.\ the Goldstone-mode effect) is included in the following 
Ansatz \cite{Wallace:1975vi} for the equation of state at low values of $x$
\begin{equation}
x\,=\,-1+a\,y^{1/2}+b\,y+c\,y^{3/2}+\cdots
\label{WZ}
\end{equation}
Note that these effects are present in $N$-vector
models along the coexistence line, i.e.\ at low temperatures and
small magnetic field (or equivalently, at low values of the variable $x$). 

The pseudo-critical line, described in the previous section,
is given by finite peaks in the susceptibility.
It characterizes the critical region when the external field is 
not zero (e.g.\ in the QCD case). The scaling form for the susceptibility along
the pseudo-critical line is given by
\begin{equation}
\chi \,=\, \partial M/\partial H \,=\, 
(1/H_0)\,{h}^{1/{\delta} - 1}
\,f_{\chi} ({t}/{h}^{1/{\beta}})\,.
\label{chisca}
\end{equation}
Note the $\chi$ has a peak at ${t_{p}}$ for each fixed ${h}$ and that
\begin{eqnarray}
{t_{p}}&=& {z_p}\,{h}^{1/{\beta} {\delta}} \\
H_0\,\chi_{p}&=& {h}^{1/{\delta} - 1} \, f_{\chi} ({z_p})\,.
\end{eqnarray}
Thus, the location of pseudo-critical line is given by $z_p$, obtained from the 
scaling function (or equation of state) for the susceptibility, which
involves the derivative of $f_M({t}/{h}^{1/{\beta} {\delta}})$.
Note also that $z_p$ is a universal constant.

For the 4-vector and the 2-vector models, the equation of state was determined 
numerically respectively in \cite{Engels:1999wf} and \cite{Engels:2000xw,Engels:2001bq},
by taking into account the
Goldstone-mode singularities and determining the location of the pseudo-critical 
line. The fitting function for $f(x)=y$ used an interpolation of two forms:
the Ansatz in Eq.\ (\ref{WZ}) at low $x$ and
Griffiths's analyticity conditions at large $x$
\begin{equation}
x(y)\,=\,A\,y^{1/\gamma}\,+\, B\,y^{(1-2\beta)/\gamma}\,+\cdots
\end{equation}

Then the equation of state $f_M({z})=M/h^{1/\delta}$ is obtained from $x(y)$.
The problem is that the ``transition'' between the two fitting forms above is close 
to the pseudo-critical point, which is itself obtained from the equation of state 
for the susceptibility [defined in Eq.\ (\ref{chisca}) above], 
involving a derivative.
It would be therefore preferable to use a parametric form for $f(x)$ without the need
for interpolation, ensuring a smooth derivative and a better determination of the
pseudo-critical point $z_p$. To this end we can consider the parametric form
\begin{eqnarray}
M &=& m_0 \, R^{\beta} \, \theta \,, \\
t &=& R \, \left( 1 - \theta^2 \right)\,,  \\
H &=& h_0 \, R^{\beta \delta} \, h(\theta) \,.
\end{eqnarray}

This yields
\begin{equation}
x = {1 - \theta^2\over \theta_0^2 - 1}\;
\left( {\theta_0\over \theta}\right)^{1/\beta}\;,\;\;\quad
f(x) = \theta ^{-\delta} \, {h(\theta)\over h(1)}\,,
\label{xf}
\end{equation}
where $h(\theta)$ is an odd function, with root given by {$\theta_0$}.
This form was introduced in \cite{Guida:1996ep} by Guida \& Zinn-Justin
for the Ising model (therefore without considering the effect due to Goldstone
modes at low $x$) and used for perturbative studies of the $N$-vector case with
the inclusion of the leading contribution of $\theta_0$.
The form leads to a smooth curve, and allows a direct relationship with critical 
amplitude ratios. We propose \cite{Cucchieri:2004xg} an improved parametrization
given by
\begin{equation}
 h(\theta) \;=\; \theta \, \left( 1 - \theta^2/\theta_0^2 \right)^2 \,
        ( 1 + \sum_{i=1}^n c_{i}\theta^{2i})\;\times
        [ \,1 + \sum_{j=1}^m d_{j}
        ( 1 - \theta^2/\theta_0^2 )^j \,]\,.
\end{equation}
The above form is based on the parametrization used perturbatively in 
\cite{Guida:1996ep} for the Ising model,
but takes into account terms associated with the effects of singularities
induced by Goldstone modes, as discussed in the Introduction. 
These new terms are included by means of the $d_j$ coefficients,
associated with an expansion around the coexistence line.
(The $d_j$'s are considered in addition to the usual $c_i$
coefficients, related to the high-temperature/high-$x$ behavior.)

\begin{figure}
\begin{center}
\ \\
\protect\vspace{-3cm}
\includegraphics[height=0.65\hsize]{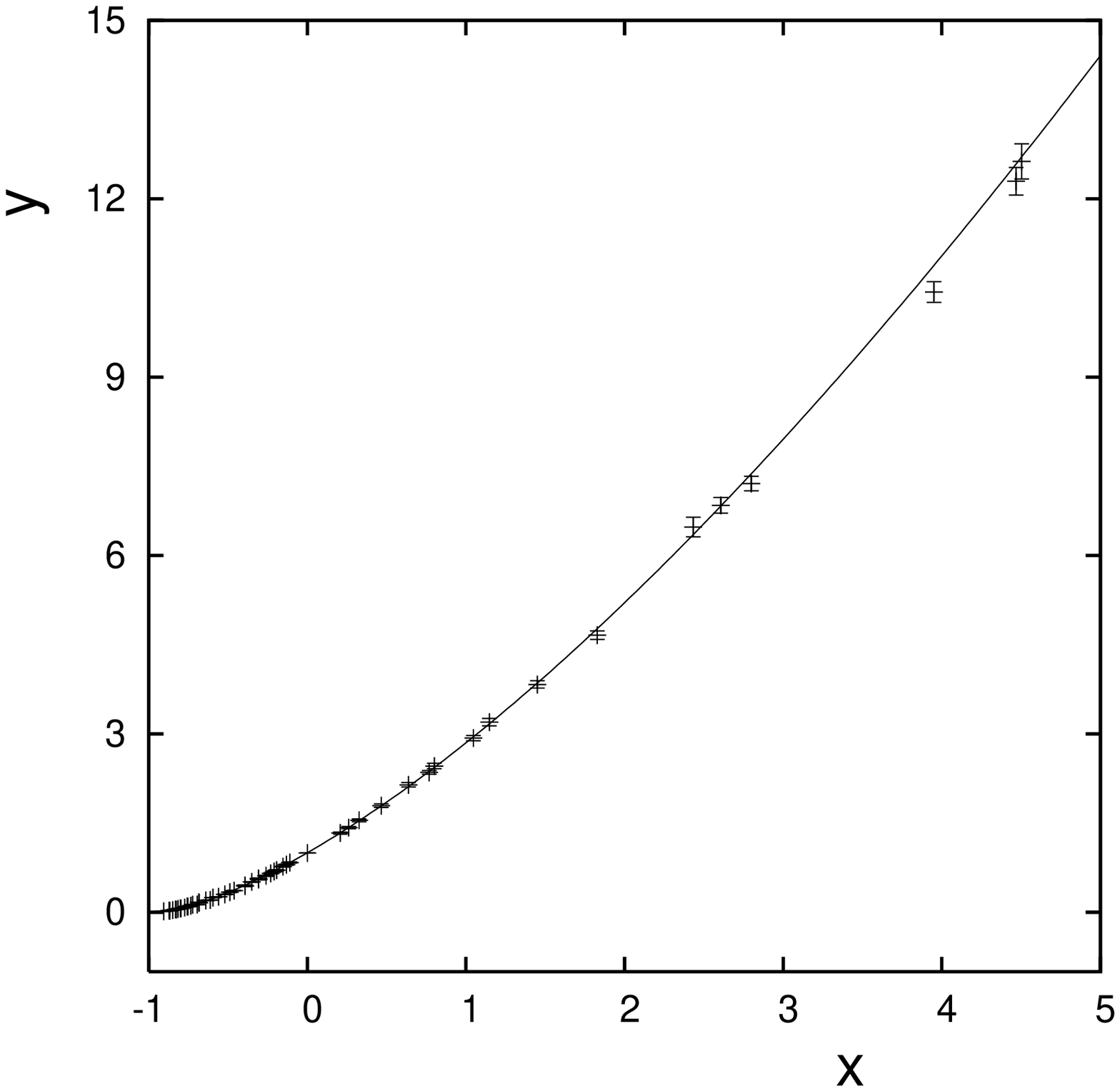}
\protect\hskip 1cm
\includegraphics[height=0.65\hsize]{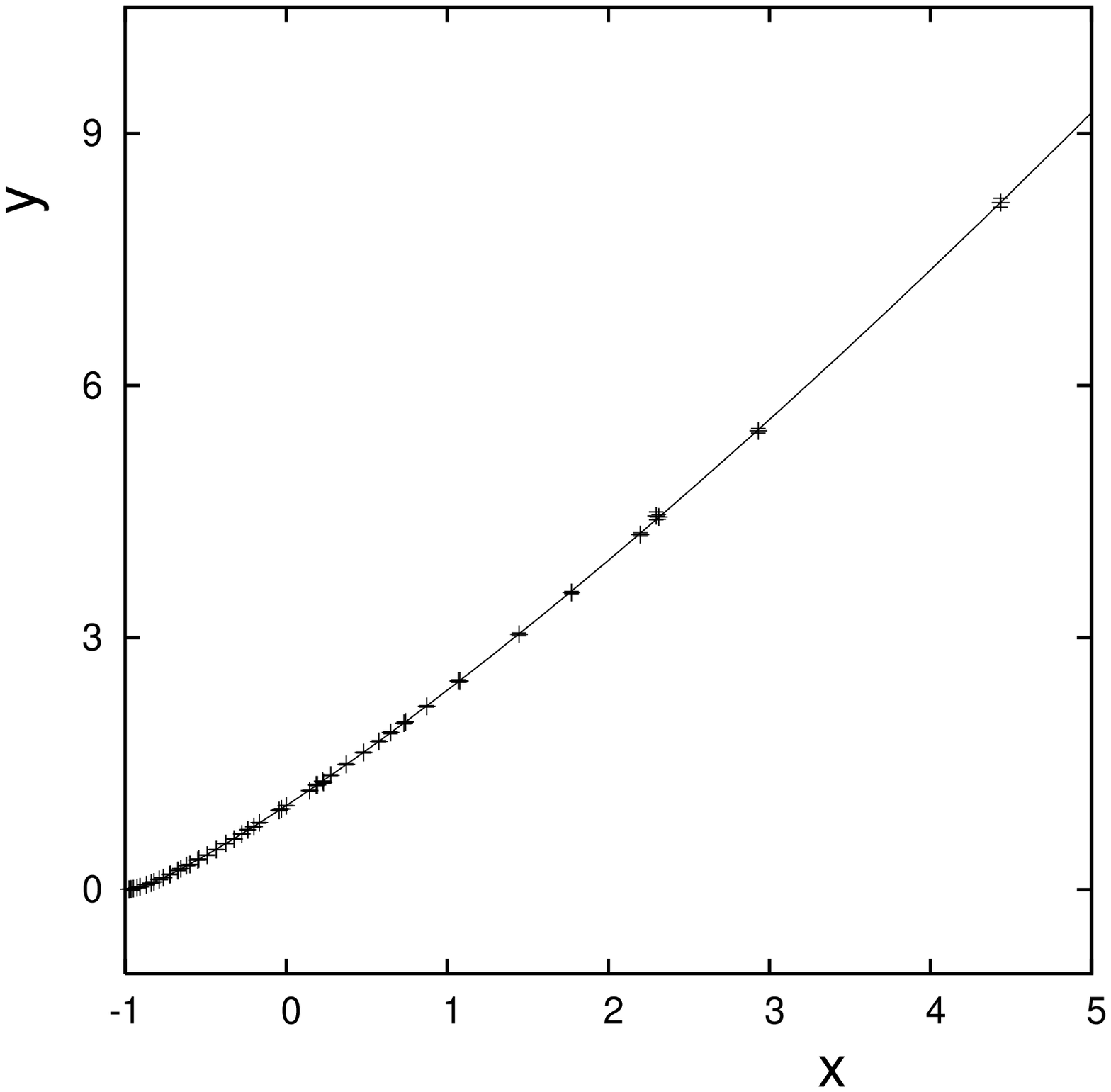}
\protect\vskip 0.4cm
\caption{Plot of the data together with the fitting form for $y(x)$
in the 4-vector case (left panel) and in the 2-vector case (right panel).
\label{fig:fit}}
\end{center}
\end{figure}

In Ref.\ \cite{Cucchieri:2004xg} we have used the proposed form above for fits 
to existing Monte Carlo data for the 4-vector case.
From our fits we see that $d_j$'s are indeed necessary for the description of 
the data. Our best fit is obtained considering (in addition to $\theta_0$)
two coefficients of type $c$ and two of type $d$
\begin{eqnarray}
\theta_0^2 &=& 2.17(4)\\
c_1 &=& 0.9(1)\,, \quad c_2 \;=\; -0.62(7) \\
d_1 &=& -1.56(4)\,, \quad d_2 = 1.15(5)\,.
\end{eqnarray}
We note that these results have errors that are one order of magnitude
smaller than the perturbative description, and fit to the data with a value of
$\chi^2$ per degree of freedom that is two orders of magnitude smaller.
The fit is shown together with the data in Fig.\ \ref{fig:fit} (left panel).
As explained above, the location of the pseudo-critical line (useful
for comparison between QCD data and the $N$-vector model's equation of state)
is obtained from the peak in the scaling function for the susceptibility
(see Eq.\ \ref{chisca}).
The peak can be determined numerically from the two equations (\ref{xf}) by 
varying $\theta$. Our result is obtained with less than 1\% of error
\begin{equation}
\theta_p \;=\; 0.587(2)\,, \;\;\qquad
z_p \;=\; 1.29(1)\,, \;\;\qquad f_\chi(z_p) \;=\; 0.341(1)\;.
\end{equation}   
The results are in agreement with previous determinations of
$z_p$ and $f_\chi(z_p)$, but our error for $z_p$ is
much smaller.

Next, we show our preliminary results for the $N=2$ case.
We have produced new data, simulating the three-dimensional
$XY$-model in the presence of a magnetic field by means of the
(Wolff) cluster algorithm. We use the improved form of the model's
Hamiltonian, introduced by Hasenbusch and T\"or\"ok for the 
zero-field case in \cite{Hasenbusch}.
We also use their values for the critical
temperature and critical exponents.
When fitting the data to our parametrization of the equation of state,
we find --- regarding the role played by the $c$ and $d$ coefficients ---
essentially the same characteristics as in the $N=4$ case
described above, with the difference that in this case 6 parameters are 
needed. Our best fit is obtained using
\begin{eqnarray}
\theta_0^2 &=& 3.25(2)\\
c_1 &=& 1.05(4)\,, \quad c_2 \;=\; -0.11(3) \,, \quad c_3 \;=\; 0.53(2)  \\
d_1 &=& -6.75(2)\,, \quad d_2 = 14.7(2)\,.
\end{eqnarray}
We show the data together with the fit in Fig.\ \ref{fig:fit} (right panel).
We see that the slope of the curve is significantly higher for the $N=4$
case, corresponding to a stronger effect of Goldstone-mode singularities,
as has already been found in \cite{Engels:2000xw,Engels:2001bq}.
Notice that the data in the $N=2$ case have smaller error bars,
leading to a very precise determination of the curve. This will enable
us to calculate (see e.g.\ \cite{Cucchieri:2004xg}) the critical amplitude 
ratio of the specific heat with the same accuracy as the experimental and
perturbative values. We quote our values for the location of the 
pseudo-critical region in the 2-vector case
\begin{equation}
\theta_p \;=\; 0.563(2)\,, \;\;\qquad
z_p \;=\; 1.61(2)\,, \;\;\qquad f_\chi(z_p) \;=\; 0.349(1)\;.
\end{equation}
As for the $N=4$ case, the results are in agreement with previous
determinations of the pseudo-critical line \cite{Engels:2000xw,Engels:2001bq}, but the
error for $z_p$ is smaller by one order of magnitude.

\section{Conclusions}

We have introduced an improved parametric form for the description
of the equation of state of $3d$ $N$-vector models. 
We show that the new parametric form indeed provides
a better fit to the numerical data as compared to previous
parametrizations. In particular, the consideration of the $d_j$
coefficients is essential for a good description of the
Monte Carlo data in the whole range of values of $x$.
Also, we were able to verify clearly the different roles played
by $c_i$ and $d_j$ parameters in the high- and low-$x$ regions.
Indeed, in this form the coefficients $c_i$ and $d_j$ contribute respectively 
to the high- 
($\theta \approx 0$) and low- ($\theta \approx \theta_0$) temperature regions.
We also stress that,
in addition to providing a better fit to the numerical data, the
expression considered is a continuous function, needing no
interpolation between the two $x$ regions. This is particularly
useful for the determination of the pseudo-critical line, since
the interpolating form introduced in \cite{Engels:1999wf} is unstable
precisely in this region. In fact, our determination of
$z_p$ is very precise in comparison to the previous estimates
from the interpolated form and the perturbative equation of state.
As a consequence of a better determination of the pseudo-critical line
in the $N=4$ case, one may get
an unambiguous normalization of QCD data for comparison to the 4-vector equation 
of state, showing better agreement for larger quark masses \cite{Mendes:2005yp,Mendes:2005wf}.
We are currently extending our analysis to the $N=2$ case. Our preliminary
results, presented here, confirm the fact that the equation of state can
be obtained with very high precision using our method. This will allow
a determination of the the critical amplitude ratio of the specific heat 
with the same accuracy as the experimental and perturbative values,
which is of a few tenths of a percent \cite{Lipa}.


\section*{Acknowledgments}

This work was supported by FAPESP and CNPq.

\end{document}